# Maximum-Likelihood Co-Channel Interference Cancellation with Power Control for Cellular OFDM Networks

Manar Mohaisen and KyungHi Chang
The Graduate School of Information Technology and Telecommunications
Inha University
Incheon, Korea
Tel: +82-32-860-8422, Fax: +82-32-865-0480
E-mail: lemanar@hotmail.com, khchang@inha.ac.kr

*Abstract*— In cellular Orthogonal Frequency Division Multiplexing (OFDM) networks, Co-Channel Interference (CCI) leads to severe degradation in the BER performance. To solve this problem, Maximum-Likelihood Estimation (MLE) CCI cancellation scheme has been proposed in the literature. MLE CCI cancellation scheme generates weighted replicas of the transmitted signals and selects replica with the smallest Euclidean distance from the received signal. When the received power of the desired and interference signals are nearly the same, the BER performance is degraded. In this paper, we propose an improved MLE CCI canceler with closed-loop Power Control (PC) scheme capable of detecting and combating against the equal received power situation at the Mobile Station (MS) receiver by using the newly introduced parameter Power Ratio (PR). At cell edge where Signal to Interferer Ratio (SIR) is considered to have average value between -5 and 10 dB, computer simulations show that the proposed closed-loop PC scheme has a gain of 7 dB at 28 km/h and about 2 dB at 120 km/h.

## I. INTRODUCTION

New generation mobile systems will provide high data rate services where good quality of communication is notably required. Orthogonal Frequency Division Multiplexing (OFDM) get the attention as one of the most promising solutions to meet these requirements [1]. In addition to its robustness against selective fading channels and narrowband interferer, OFDM system has a lower implementation complexity compared with single carrier systems with equalizer or RAKE receiver [2], [3].

When OFDM is applied to cellular communication systems, frequency reuse is needed to increase the system capacity and to ameliorate the frequency utilization of the network. On the other hand, frequency reuse in cellular systems causes Co-Channel Interference (CCI) which is one of the major factors that limit the capacity of the system [4]. In order to reduce the effect of CCI in OFDM networks, CCI cancelers based on Maximum Likelihood Estimation (MLE) have been proposed in [5] and [6].

The MLE scheme generates replica signals for desired and CCI signals from all possible weighted combinations of the desired and CCI signals. The weights represent the estimated channel transfer functions at the sub-carriers. The replica that has the smallest Euclidean distance from the received signal is selected and data are detected [5]. When the received power of the desired and the interference signals is nearly the same, several signals combinations result in similar replicas that give the same Euclidean distance from the received signal. As consequence, the MLE cannot distinguish between signals and the BER performance is degraded [6].

To solve this problem, Frequency Interleaving (FI) and Frequency Spread Coding (FSC) can be applied [6], [7]. The BER performance improvement by applying FI depends on the interleaver depth while the improvement using FSC increases as the interval between sub-carriers holding the same data symbol increases. The performances of FSC and FI schemes are mainly dependent on the selectivity of the channel; if the frequency selectivity of the channel is not significant then the BER performance improvement becomes negligible. In addition, the complexity of the MS receiver increases[6].

In this paper, we propose an improved MLE CCI cancellation by applying closed-loop power control (PC) scheme to combat against the situation where received power of desired and interference signals is nearly the same. For two-cell scenario, the newly introduced average signal to interference Power Ratio (PR) is calculated for frame $k$. If the PR is less than a certain threshold, Mobile Station (MS) sends a feedback to the Power Control Unit (PCU) at the serving Base Station (BS) where the PCU boosts the transmission power for the frame $k+1$. The proposed closed-loop PC scheme significantly reduces the degradation of BER performance caused by the CCI. This two-cell scenario can be extended to more than one interferer signals scenario.

This paper is organized as follows: CCI effects on the performance degradation in cellular OFDM networks and BER performance of MLE canceler are presented in section 2. In section 3, we describe the proposed MLE CCI canceler with closed-loop PC scheme by presenting the BS transmitter and MS receiver structures in detail. Simulation results for different mobilities and under different channel environments are presented in section 4. Finally, we draw conclusions in section 5.

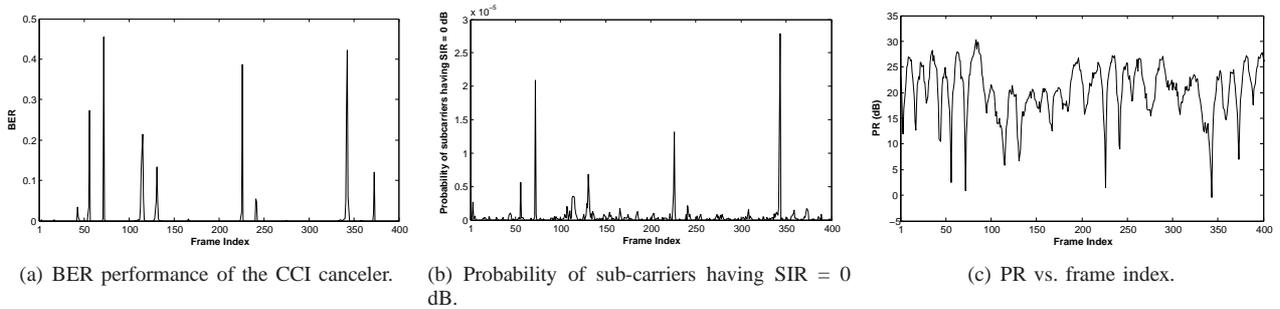

(a) BER performance of the CCI canceler.  (b) Probability of sub-carriers having SIR = 0 dB.  (c) PR vs. frame index.

Fig. 1. BER performance behavior of MLE CCI canceler.

## II. CCI IN CELLULAR OFDM NETWORK

### A. Cellular OFDM Networks with CCI

When OFDM is applied to cellular networks, frequency reuse is used to increase the spectral efficiency of the overall system. By applying frequency reuse, different BS geographically separated by a reuse distance use the same frequency. This leads to a mutual CCI among BS. CCI degrades the BER performance at MS served by these BS and in severe situations it leads to a link failure between MS and serving BS [8]. In the following subsection, we present the theoretical effect of CCI on the BER degradation.

### B. Effect of CCI on the BER Performance Degradation in OFDM Cellular Networks

Herein, we present a general formula of the effect of CCI on the BER performance degradation in OFDM cellular networks.

After some simplifications and derivation based on results presented in [9], the BER performance in OFDM cellular network with CCI for QAM modulation is given by

$$P_e = \frac{1}{log_2(\sqrt{M})} \cdot \left(1 - \frac{1}{\sqrt{M}}\right) \cdot \left[1 - \frac{1}{\sqrt{\frac{M-1}{3}\left[\sum_{j=2}^{K+1}\frac{1}{SIR_j} + \frac{2}{(log_2(M))\frac{E_b}{N_0}}\right] + 1}}\right] \quad (1)$$

where $M$ is the modulation size, $K$ is the number of CCI interferer BS, $SIR$ is the signal-to-interferer ratio and $E_b/N_0$ is the average signal-to-noise ratio.

### C. BER Performance Behavior of MLE CCI Canceler

When the received power of the desired and interference signals is nearly the same (i.e. SIR $\cong$ 0 dB), the BER performance is degraded, because there are several combinations that result in the same replicas. This leads to ambiguity at the MS receiver because several replicas have the same minimum Euclidean distance.

Fig. 1 (a) and (b) show an example of the BER performance of the MLE canceler and the probability of sub-carriers having SIR = 0 dB where a single CCI signal is considered, respectively. The BER performance is degraded when the probability of sub-carriers having 0 dB increases. In practice, the change in the instantaneous SIR is very rapid. This is why we introduce the new term Power Ratio (PR) whcih has slower change. For two-cell scenario, PR is given by

$$PR = 10 \cdot log_{10}\left(\frac{\sum_{i=1}^{N}\hat{h}_{1i}^2}{\sum_{i=1}^{N}\hat{h}_{2i}^2}\right) \quad (2)$$

where $\hat{h}_{1l}$ and $\hat{h}_{2l}$ are the estimated channel transfer functions between MS and desired and interferer BS at at the $l^{th}$ sub-carrier, respectively. Fig 1 (c) shows the PR values versus frame index. As it is clear from this figure, the PR is inversely proportional to the probability of sub-carriers having SIR = 0 dB; when the PR value decreases under a certain threshold, the probability of sub-carriers having SIR = 0 dB increases and consequently the BER performance is degraded. Therefore, the PR value can be used to indicate the probability of sub-carriers having SIR = 0 dB.

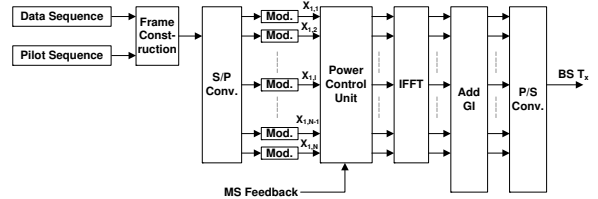

Fig. 2. BS transmitter structure.

In the next section, the proposed MLE CCI canceler with closed-loop PC is explained in more detail.

## III. MLE CCI CANCELER WITH PC

Fig. 2 shows the BS transmitter structure. After frame construction, data are serial to parallel converted to $N$ parallel bit streams. Each bit stream is modulated using QPSK modulator. If the MS feedback is 1, the power control unit boosts the transmission power of next frame by 3 dB otherwise transmission power is not boosted. The output of the power control unit is fed to the IFFT block to get the time domain signal. Finally, the guard interval (GI) is added and data are parallel to serial converted and then transmitted.

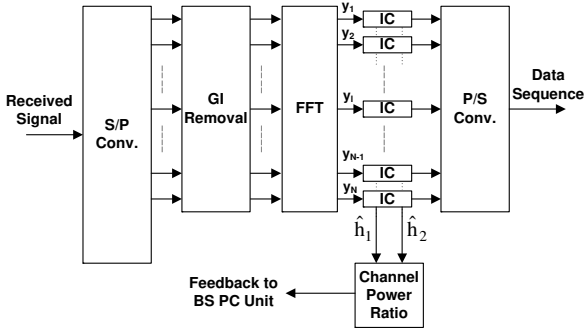

Fig. 3. MS receiver structure.

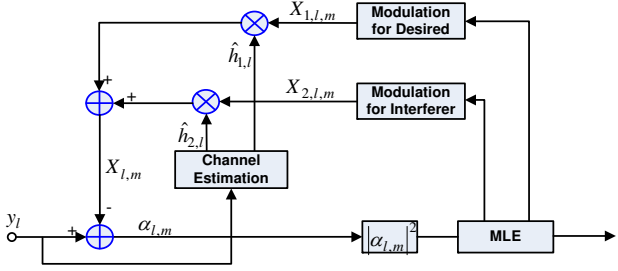

Fig. 4. CCI Canceler.

Fig. 3 shows the MS receiver structure. At first, received signal is serial to parallel converted then the GI is removed. The FFT is applied to get the frequency domain signal $Y$.

The received signal on the the $l^{th}$ sub-carrier is given by

$$y_l = \sum_{k=1}^{K+1} h_{k,l} x_{k,l} + n_l \quad (3)$$

where $x_{k,l}$ is the symbol transmitted on the $l^{th}$ sub-carrier by the $k^{th}$ transmitter and $n_l$ is filtered AWGN on the $l^{th}$ sub-carrier.

Fig. 4 shows the CCI Canceler (IC) for $K = 1$ at the $l^{th}$ sub-carrier. The MLE block outputs a set of desired and CCI candidates which are modulated and weighted using the estimated channel transfer functions and added to generate the replica which is indexed by $m$. Each generated replica is compared to the received signal on the $l^{th}$ sub-carrier. The resulting Euclidean distance between the replica $m$ and the received signal $y_l$ is given by

$$\alpha_{l,m} = y_l - \sum_{k=1}^{2} \hat{h}_{k,l} x_{k,l,m} + n_l. \quad (4)$$

The MLE block chooses the candidate of the desired and CCI which has the least $|\alpha_{l,m}|^2$.

The estimated channels transfer functions between MS and the desired and CCI BS ($\hat{h}_1$ and $\hat{h}_2$) are fed to the channel PR unit which calculates the PR value and compares it with a fixed threshold dependent on the mobile speed and average SIR value. If the PR is less than the threshold, the MS sends a feedback to the desired BS's PCU to boost the transmission power for the next frame.

TABLE I

PR THRESHOLD VALUES FOR DIFFERENT SIR VALUES AND MS SPEEDS.

| SIR (dB) \ Speed | 10 km/h | 28 km/h | 120 km/h |
|---|---|---|---|
| -20 | -20 | -20 | -15 |
| -15 | -15 | -15 | -10 |
| -10 | -10 | -7 | -7 |
| -5 | 2 | -4 | -4 |
| 0 | 2 | 0 | 0 |
| 5 | 5 | 5 | 4 |
| 10 | 10 | 10 | 10 |
| 15 | 12 | 15 | 15 |
| 20 | 14 | 18 | 18 |
| 25 | 17 | 20 | 22 |
| 30 | 20 | 22 | 24 |
| 35 | 22 | 24 | 26 |

## IV. SIMULATION RESULTS

Computer simulations were performed to examine the performance of the MLE with and without closed-loop PC to show the gain of the proposed algorithm. Table I shows threshold lookup table (in dB) used at MS for different average SIR values at different mobilities. These threshold values are adjusted to get best CCI cancellation gain with lowest power loss. Table II shows the principal parameters used to perform these simulations. The number of CCI signals $K = 1$; this means that a single CCI signal was considered. An OFDM frame has 4 pilot symbols used for channel estimation and 51 data symbols. The overall OFDM frame duration is 220 $\mu$s. The desired BS pilots are inserted on the odd indexed sub-carriers of the odd indexed pilot OFDM symbols and on the even sub-carriers of the even indexed pilot OFDM symbols. While the CCI BS pilots are inserted on the even indexed sub-carriers of the odd indexed pilot OFDM symbols and on the odd indexed sub-carriers of the even indexed OFDM symbols. In the frequency domain, the channels transfer functions are estimated by averaging over the four pilot symbols in the time domain. When the PR decreases under the specified threshold, MS sends a feedback to the desired BS asking it to boost the power by 3 dB for the next frame to be transmitted. The threshold value is dependent on the average SIR value and on the MS speed. We assume that the MS can measure the average SIR and its speed which can be obtained from the maximum Doppler shift. To detect the transmitted symbol at every sub-carrier, $M^K$ replicas are calculated where $M$ is the

TABLE II

SIMULATION PARAMETERS.

| Parameter | Value |
|---|---|
| Carrier Frequency | 2 GHz |
| Number of Cells | 2 |
| Bandwidth | 20 MHz |
| FFT Size | 64 |
| Guard Interval | 16 |
| Modulation | QPSK |
| Speed | 10, 28 and 120 km/h |

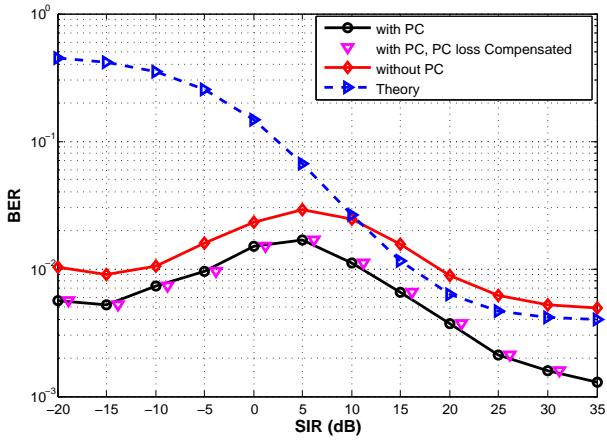

Fig. 5. BER under single-path fading channel; speed = 10 km/h and $E_b/N_0$ = 18 dB.

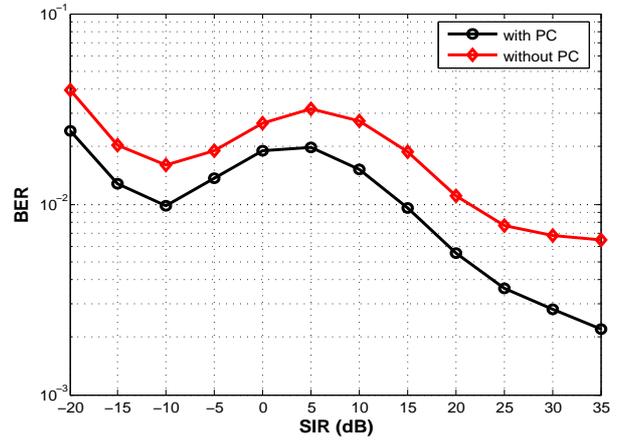

Fig. 7. BER under single-path fading channel; speed = 28 km/h and $E_b/N_0$ = 18 dB.

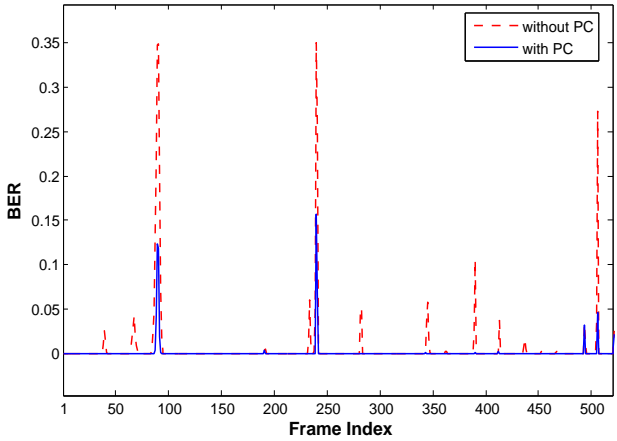

Fig. 6. BER performance improvement by closed-loop PC.

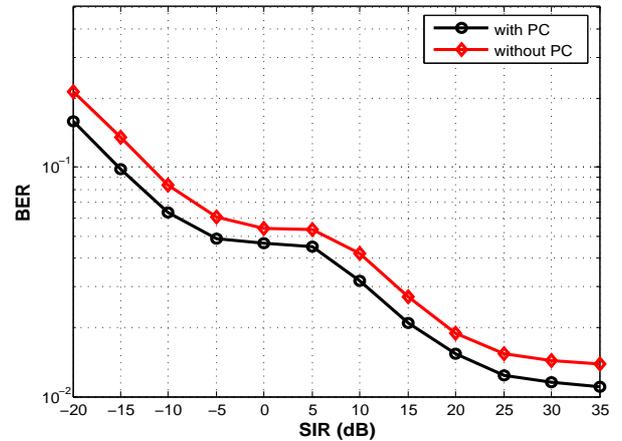

Fig. 8. BER under single-path fading channel at 120 km/h.

modulation size and $K$ is the number of BS. For example, for QPSK modulation ($M = 4$) and single CCI signal, the MS receiver should calculate $4^2 = 16$ replicas to detect the transmitted symbol. This indicates that the overall complexity of the proposed algorithm is acceptable compared to other CCI cancellation schemes. Furthermore, applying the proposed CCI cancellation scheme required a feedback of only one bit which can be guaranteed with significant accuracy at low and medium mobilities.

Fig. 5 shows the BER performance of the MLE CCI cancellation scheme with and without the proposed closed-loop PC scheme under signal-path fading environment. In addition, the theoretical BER performance without any CCI cancellation given in section II is shown as dashed line. The MS speed and average $E_b/N_0$ are set to 10 km/h and 18 dB, respectively. After taking the closed-loop PC power loss into consideration, the proposed closed-loop PC scheme has a gain of about 7.5 dB for users at the cell edge. Users at the cell edge are considered to have an average SIR value between -5 and 10 dB.

Fig. 6 shows the BER performance of the MLE CCI cancellation algorithm with and without the proposed closed-loop PC scheme. The PR is used to detect the probability of sub-carriers having SIR = 0 dB where the biggest part of the BER occurs. The dashed line curve represents the BER of the MLE CCI cancellation scheme without closed-loop PC while the solid line represents the BER with closed-loop PC scheme applied. As shown, applying the closed-loop PC scheme leads to important decrease in the BER and some frames become error free.

Fig. 7 shows the BER performance of the MLE CCI cancellation algorithm with and without the proposed closed-loop PC scheme. Average $E_b/N_0$ is set to 18 dB and mobility to 28 km/h. The Doppler frequency is 52 Hz at carrier frequency 2 GHz. The gain of the proposed closed-loop PC scheme is 6 dB after taking the closed-loop PC power loss into consideration.

Fig. 8 shows the BER performance of the proposed closed-loop PC scheme for average $E_b/N_0$ set to 18 dB at 120 km/h. Even at this high mobility, the MLE CCI canceler with the proposed closed-loop PC scheme gives a 2 dB of gain compared to the MLE CCI canceler without PC.

Fig. 9 shows the BER performance under two-path fading channel environment. The delay between the two paths is set to 6 samples corresponding to 300 ns and the relative power profile of the second path is -8 dB. For users at the cell edge, who are the target for any CCI cancellation algorithm, the gain is about 7 dB.

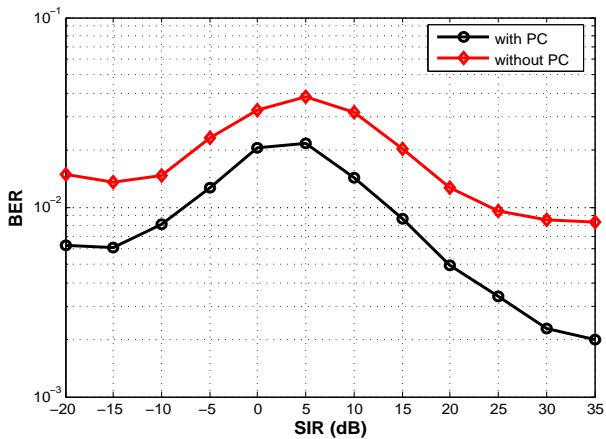

Fig. 9. BER under two-path fading channel; speed = 10 km/h and $E_b/N_0$ = 18 dB.

## V. Conclusions

To reduce the degradation in the BER performance in cellular OFDM networks, MLE CCI canceler has been proposed in the literature. When the received power of the desired and CCI signals is nearly the same, different replicas result in minimum Euclidean distance from the received signal and therefore the BER performance of the MLE CCI canceler is degraded.

In this paper, we propose a closed-loop PC scheme capable of detecting the situation where equal power from desired and CCI BS is received. To detect this situation, we introduced the new PR parameter which represents the ratio between the average powers of estimated channels of the desired and CCI BS.

Computer simulations were performed to investigate the performance improvement of the proposed closed-loop PC scheme. The closed-loop PC gain was calculated for users at cell edge where average SIR is considered to have values between -5 and 10 dB. At low mobility of 10 to 30 km/h, the gain of the proposed closed-loop PC scheme is about 6.5 dB under single-path and 2-path fading channels. While for high mobility of 120 km/h, the gain of the proposed closed-loop PC scheme is about 2 dB in the average SIR. These results show that the proposed closed-loop PC scheme shows excellent BER performance improvement and this significantly increases the cellular OFDM system capacity.